\documentclass[usegraphicx,useAMS,usenatbib]{mn2e}
\usepackage{graphicx}

\title{A Model of Solar Dynamo with Alternative Conversion of Large-Scale Magnetic Field and Production of Sunspots}

\author[Biping Gong]
    {Biping Gong \\
Department of Physics, Huazhong University of Science and
Technology, Wuhan 430074, China  \\
}

\begin{document}

\date{Accepted, Received ;}

\pagerange{\pageref{firstpage}--\pageref{lastpage}} \pubyear{2017}

\maketitle

\label{firstpage}

\begin{abstract}
{Since the discovery of  solar cycle related with magnetic field in 1908, deep seated oscillatory dynamo  has been studied extensively. 
However,  there are still open questions on the solar dynamo, e.g.,  asymmetric conversion  between  large-scale poloidal and toroidal field  as well as  physics underlying the butterfly pattern of sunspots.    
Here we report a new generation of large-scale magnetic field and process of energy release.
The inductive action of fluid motions pervading the solar interior is represented by a RLC circuit in which the toroidal field built up through twisting of poloidal field, so called $\omega$-effect,  plays the role of  a capacitor.
 Such a RLC circuit not only provides a self-sustained oscillatory system avoiding Cowling's antidynamo theorem, but also site of rapid magnetic reconnection which reproduces quadrupole magnetic field  interpreting the behavior of sunspots and moving of foot-point in solar activities. 
Moreover, parameters of the circuit and the Sun  are well consistent with the 22-year solar cycle. }
\end{abstract}


\section{Introduction}





George Ellery Hale first linked magnetic fields and sunspots in 1908~\citep{Hale08} , who proposed  that the sunspot cycle period is 22 years, covering two periods of increased and decreased sunspot numbers, accompanied by polar reversals of the solar magnetic dipole field. 
It indicated the existence of toroidal field, residing in solar interior as the source of sunspots 

In 1919 Larmor proposed the inductive action of fluid motion as origin of such  magnetic field, from which
the twisting of large scale poloidal magnetic field by differential rotation in the solar interior 
 is responsible for equatorial antisymmetry of the solar internal toroidal field.

However, in 1933 Cowling demonstrated that   even the most general purely axisymmetric flow could not sustain an axisymmetric magnetic field against Ohmic dissipation by themselves,  which is known as Cowling's antidynamo theorem.     

The theorem works  e.g., in a Faraday's disk, consisting of an electromotive force (emf)  of the disk, $E$, a power, 
${\bf u}\times {\bf B}=u_{\phi}B_z$; and a term of energy dissipation, $\eta J$, 
\begin{equation}{\label{ohm0}}
{\bf E}=\eta {\bf J}-{\bf u}\times {\bf B}
\end{equation}
Where $J$ denotes current density, $u_{\phi}$ and $B_z$ represents plasma velocity due to the rotation of the disk, and poloidal field respectively as shown in Fig.$\ref{disk}$.
In fact,  Eq.$\ref{ohm0}$ is  equivalent to a RL circuit which cannot sustain by itself.
 




In 1950s Parker suggested  that  cyclonic twist could be introduced by Coriolis force which gives rise turbulent fluid elements in the solar convection zone, and provides a way of   conversion from toroidal field to poloidal field ($\alpha$ effect).  This is equivalent to an  additional term to   the right hand side of  Eq.$\ref{ohm0}$ which can make  a self-sustained system  and avoid   Cowling's theorem. This idea inspired subsequent development of mean-field electrodynamic, which  became the mainstream solar dynamo models.

However, there are still some problems in the  mean-field electrodynamics. E.g., 1) internal differential rotation required by the  mean-field models deviates from the result of  helioseismology. 

2) In the context of  buoyancy, magnetic fields strong enough to produce sunspots could not be stored in the solar convection zone for sufficient lengths of time for adequate amplification.

3) $\alpha$ effect and magnetic diffusivity operating in the mean-field electrodynamics was also called into question by theoretical calculations and numerical simulations \citep{PC10}.

In this paper, an additional term denoting the toroidal  magnetic field generated through stretching of poloidal field through differential rotation of near the solar photosphere is added to the RL circuit of Faraday disk, as shown in  Eq.$\ref{ohm0}$. Then    
 a self-excited RLC circuit is obtained, in which the toroidal field  is equivalent to a capacitor originating in the $\omega$ effect.  
 
 The role of $L$, $C$ in the RLC circuit, their derivation and relationship with solar parameters   have been discussed in  this paper. These issues have not been addressed concretely although RLC solar  cycle model has been discussed before\citep{Poly96}.

 In the new model, the conversion between the polar and toroidal field is through the  emf of the RLC circuit, which is symmetric and  easy to keep equivalence time of conversion between them (Section 2).





With the $\omega$ effect which build up toroidal field through stretching of poloidal field via differential rotation of the Sun, the  toroidal field of prior solar cycle and the current one of opposite polarity are concentrated at lower and lower latitude, so that a X-point configuration is formed.  
 Such a configuration together with radial emf produced by the RLC circuit provide sites of rapid magnetic reconnection,  in which magnetic field reproducing sunspots originates in Hall current rather than  buoyancy of magnetic field (Section 3).
How such a modified  Faraday disk model confronts with Cowling's antidynamo theorem is addressed (Section 4).   
A number of questions concerning solar cycle can be explained  under such a new dynamo model (Section 5). 

Consequently, the generation of large-scale field, the site and configuration of energy release of magnetic field, and mechanism of reproducing sunspots of this paper all differ from previous ones.

\section{The twisted magnetic field and the RLC circuit}

 In the scenario of a Faraday disk, 
the first term at right hand side of Eq.$\ref{ohm0}$  corresponds to an emf, $E_r=u_{\phi}B_z$. Integrating this emf along the convection zone  of the Sun is read, 
 \begin{equation}{\label{Er}}
\xi_r=\int_{r_2}^{r_1} E_r dr=\frac{\Omega }{2}(r_2^2-r_1^2)B_z
\end{equation}
where  $r_1=0.95R_{\odot}$ and $r_2=0.74R_{\odot}$  (where  $R_{\odot}$ is the radius of the Sun). 




 The azimuth motion of plasma around the axis of the Faraday disk, $u_{\phi}$, as shown in Fig.$\ref{disk}$, corresponds to an inductance of,
\begin{equation}{\label{Bflux}}
L\approx\frac{\mu}{2r_1}\pi(r_1^2-r_2^2)=5.2 \times 10^2 \, (H)
\end{equation}
where $\mu=\mu_0$  is the permeability, and the area of the disk is of $S\approx\pi(r_1^2-r_2^2)= 5.5\times 10^{17} m^2$.

The left hand side of  Eq.$\ref{ohm0}$ corresponds to an emf of, 
 \begin{equation}{\label{EP}}
 \xi_P=-S\frac{\partial B_z}{\partial t} =-L\frac{\partial I}{\partial t}
 \end{equation} 
 where $I$ is the current of the circuit. 
 
   
 The second term at right hand side of  Eq.$\ref{ohm0}$ corresponds to energy dissipation of the disk,  
   $\eta \int^{r_1}_{r_2}jdr=\eta\frac{aI}{\Sigma}=RI$, where $\Sigma$ is the cross-section of the current in the disk and $a=r_1-r_2$.
 With the three terms above,  Eq.$\ref{ohm0}$ becomes, 
\begin{equation}{\label{RL}}
\xi_r-L\frac{\partial I}{\partial t}-RI=0 
\end{equation}
Apparently,  such a circuit cannot self-sustained.

In the case of the Sun, the  twisted field magnetic field, $B_{\phi} $ also induces electromotive force in radial direction  in the right hand side of  Eq.$\ref{ohm0}$. Taking into account of it,  ${\bf E}_R={\bf u}^{\prime}\times {\bf B}^{\prime}$,   the general Ohm's law becomes,  
\begin{equation}{\label{ohm1}}
{\bf E}=\eta{\bf J}-{\bf u}^{\prime}\times {\bf B}^{\prime}-{\bf u}\times {\bf B} 
\end{equation}
As indicated by  helioseismology \citep{Tom03},  different latitude corresponds to  different rotation speed. E.g., at latitude of $60^{\circ}$ the rotation speed at 0.95$R_{\odot}$ is only  380nHz, which is  much slower than that at the equator (470nHz). And at latitude $15^{\circ}$, $30^{\circ}$, and $45^{\circ}$ at radius 0.95$R_{\odot}$ the rotating speeds are between 380nHz and 470nHz. 
Considering polar field line, $B_z$, at radius  0.95$R_{\odot}$  is frozen  between  latitude $60^{\circ}$ and  the equator, then  the deviation in differential rotation speed between the two latitudes is of  100nHz\citep{Tom03}.  In such a discrepancy,  the field line is approximately  wounded up for  3 times per year. 

In contrast, by the helioseismology \citep{Tom03}, the  rotation speed 
are of 470nHz and 460nHz  at $r_1$ and $r_2$ respectively in the equatorial plane.  In other words, differential rotation is much significant in latitude than that in radial, so that  the  differential rotation in radial ($r_1$ and $r_2$) is neglected, as shown in Eq.$\ref{Er}$, which is treated as having the same rotation velocity, $\Omega$.


\cite{LB96} defined a pitch of between the poloidal and wound up toroidal field.
The twisting field can be written as, 
\begin{equation}{\label{twist}}
B_{\phi}=\kappa\int ^t_0 B_z dt
\end{equation}
where $\kappa\equiv{n}/{P_0}$, 
with $n$ the number of twisting and $P_0$  the time scale of twisting field lines.
In the case of $P_0\approx 1$years, and $n=3$, 
and with polar field  of $B_z\approx2$G, 
the toroidal field  $B_{\phi}$ can be magnified up to $B\approx 30$G  through winding of field line in 5 years.
Apparently,  the wound up toroidal field, $B_{\phi}$,  lags the polar field $B_z$.
Hence,  the general Ohm's law,
 Eq.$\ref{ohm1}$ can be written as, 
 \begin{equation}{\label{ohm2}}
E_P=\eta J-u_{\phi}B_{z}+u_z B_{\phi}
\end{equation}

The polar flux of the Faraday disk varies with $SB_z=LI=LI_0\sin\omega t$, so that $B_z \propto I$. In comparison, 
 Eq.$\ref{twist}$ indicates $B_{\phi}\propto \int^t_0 B_zdt \propto \int^t_0 Idt \propto q$.  Therefore, the role of $B_{\phi}$ in is equivalent to a capacitor in an RLC circuit,
\begin{equation}{\label{qc}}
\xi_R=\int^{r_1}_{r_2} E_Rdr=\int^{r_1}_{r_2} u_zB_{\phi}dr=u_zB_{\phi}a=\frac{q}{C}
\end{equation} 
Time derivative of Eq.$\ref{qc}$ and using Eq.$\ref{twist}$ yields, 
\begin{equation}{\label{CB}}
\frac{1}{C}\frac{d q}{dt}=\frac{1}{C}I={\kappa u_z} B_za=\kappa u_z\frac{\mu Ia}{2r_1}
\end{equation}  
With $u_z\approx 25m/s$, $\kappa\approx 3/P_0$, $\mu=\mu_0$, and $r_1=0.95R_{\odot}$, the equivalent capacity is read, 
\begin{equation}{\label{C}}
 C=\frac{2r_1}{\mu\kappa u_z a}\approx 2.1\times 10^{13} (F) . 
\end{equation} 

On the other hand, with a height of a cylinder of  $\Delta h= R_{\odot}$, the capacity of a conventional capacitor , $C^{\prime}$, can be calculated. Assuming $C=C^{\prime}$ we have $\epsilon=1.2\times 10^3 (F/m)$ as shown in Table~1.



 The time derivative of general Ohm's law  of Eq.$\ref{ohm2}$ gives, 
 \begin{equation}{\label{RLC1}}
L\frac{d^2I}{dt^2}+\frac{I}{C}+R\frac{dI}{dt}=\frac{\partial \xi_r}{\partial t}
\end{equation}


  
 




With the values of $L$ and $C$ given by  Eq.$\ref{Bflux}$ and Eq.$\ref{CB}$, the oscillation period of such a RLC circuit as given by Eq.$\ref{RLC1}$ is read, 
\begin{equation}{\label{22c}}
P=2\pi\sqrt{LC}=2\pi\sqrt{5.2\times 10^2 \times 2.1\times 10^{13}}\approx 21 (year)
\end{equation}
 In other words,  the RLC circuit with the capacitor and inductance derived from the parameters such as $u_z$, $\kappa$, $S$, $r_1$, and $\mu$, $\xi$ automatically oscillates at a period of approximately 22-year.





The impedance of  the resistor, $R$, the coil, $L$, and the capacitor, $C$, in the circuit of Eq.$\ref{RLC1}$   are
$Z_R=R$,   $Z_L=i\omega L$,  $Z_C=\frac{1}{i\omega C}$,  respectively. 
The total impedance of such a RLC circuit is 
 
\begin{equation}{\label{impedance}}
 Z=\sqrt{(\omega L-1/\omega C)^2+R^2} 
 \end{equation}
 The power of the circuit totally  dissipated on the resistor requires that $Z=R$ be satisfied in Eq.$\ref{impedance}$, which means $\omega L=1/\omega C$. This is  consistent with 
 Eq.$\ref{22c}$. 
 



The polar field $B_z$ and the twisted field $B_{\phi}$ correspond to the first (L) and second term (C) of   Eq.$\ref{RLC1}$ respectively, which can converse each other at period given by Eq.$\ref{22c}$. Therefore, in the new dynamo the equal time of conversion between the polar and twisted field is a natural result of the RLC circuit.

The closed lines  of Fig.$\ref{disk}$A represent a unit of RLC circuit, infinite number of which form the equivalent circuit as shown in the middle of  Fig.$\ref{disk}$.  Such a RLC circuit  can be described by Eq.$\ref{RLC1}$ with an intrinsic frequency of oscillation of  $\omega_0=2\pi/\sqrt{LC}$ as shown in Eq.$\ref{22c}$.

At the right hand side of  Eq.$\ref{RLC1}$  is a derivative of $\xi_r$ given by Eq.$\ref{Er}$, which can be rewritten  as 
\begin{equation}{\label{ErT}}
\frac{\partial \xi_r}{\partial t}=b\frac{\partial B_z}{\partial t}
\end{equation}
where $b=\Omega(r_2^2-r_1^2)/2$.
Substitute  $B_{z}\approx\mu I/2r_1$, and  the current of the RLC circuit of $I=I_0\sin\omega t$, into   Eq.$\ref{ErT}$, we get,  
\begin{equation}{\label{ErT1}}
\frac{\partial \xi_r}{\partial t}=\frac{\omega b\mu I_0\cos\omega t}{2r_1}, 
\end{equation}


The oscillation of the RLC circuit of  Eq.$\ref{RLC1}$ can be seen as a response of the circuit to the external ``force", 
${\partial \xi_r}/{\partial t}$   of Eq.$\ref{ErT1}$, 
\begin{equation}{\label{amp}}
I= \frac{h}{[(\omega_0^2-\omega^2)^2-4\gamma^2\omega^2]^{1/2}}\cos[\omega t+\alpha]
\end{equation}
where $h=\omega b\mu I_0/(2r_1)$, $\gamma={R}/{(2L)}$, and $\beta=\alpha+\frac{\pi}{2}=tan^{-1}\frac{\omega_0^2-\omega^2}{2\gamma\omega}$.

%

As shown Fig.$\ref{spot}$, the phase  discrepancy between the  $\sin$ function of the polar field, and $\cos$ function denoting sunspot number is of  $\delta=9.7^{\circ}$,  which requires the angle of $\beta$ defined under Eq.$\ref{amp}$ to be  $\beta=99.7^{\circ}$. 

To have such a $\beta$ in the case of  $|\omega-\omega_0|/\omega_0\sim10^{-1}$, demands  
\begin{equation}{\label{tan}}
tan^{-1}[\frac{-\omega_0}{10\gamma}]\approx 99.7^{\circ}
\end{equation}
which requires, $\gamma\approx 1.5\times 10^{-10} $.  In such a case we have,  $R=2L\gamma\approx 1.5\times10^{-7} (\Omega)$ (so that $\sigma=7\times10^6 S/m$). 







Recall the flux of  polar field varies with $SB_z=LI=LI_0\sin\omega t$, the phase of which deviates from that of  energy dissipation of  $I^2R\propto \cos^2[\omega t+9.7^{\circ}]$ for approximately $\pi/2$. 
 
In contrast, the change of $B_{\phi}$ lags   about $\pi/2$ from the polar field due to the effect of twisting as shown in Eq.$\ref{twist}$, so that 
the twisted field varies with   $B_{\phi}\propto \cos\omega t$.
Therefore, the energy of the twisted field varies with  $\frac{B_{\phi}^2}{2\mu}V\propto\cos^2\omega t$;  which is approximately in phase 
with $I^2R\propto\cos^2(\omega t+9.7^{\circ})$. 



The energy stored in the twisted field,$\frac{B_{\phi}^2}{2\mu}V$   is difficult to estimate due to the uncertainty in the involved volume $V$.
Alternatively, it can be estimated  by the equivalent current in the term, 
$B_{\phi}\approx \mu I_z/(2\pi r_1)$. With $B_{\phi}\approx 1\times10^{-3}T$, we have $I_z\approx 3\times 10^{12} A$, corresponding to a power of $I_z^2R\approx 1.4\times 10^{18} W$,  so that it can dissipate the  energy of  $1\times 10^{25}$J in every 5 months. 

Then in a half solar cycle of 11 years, the energy release from such a RLC circuit is of $\sim 10^{26}$J. Whereas, such an amount of energy is not released evenly in 11 years, instead the majority energy is released in a few years during the peak of the toroidal field through magnetic reconnection.





\section{Magnetic reconnection}

\begin{figure}
\includegraphics[width=0.45\textwidth]{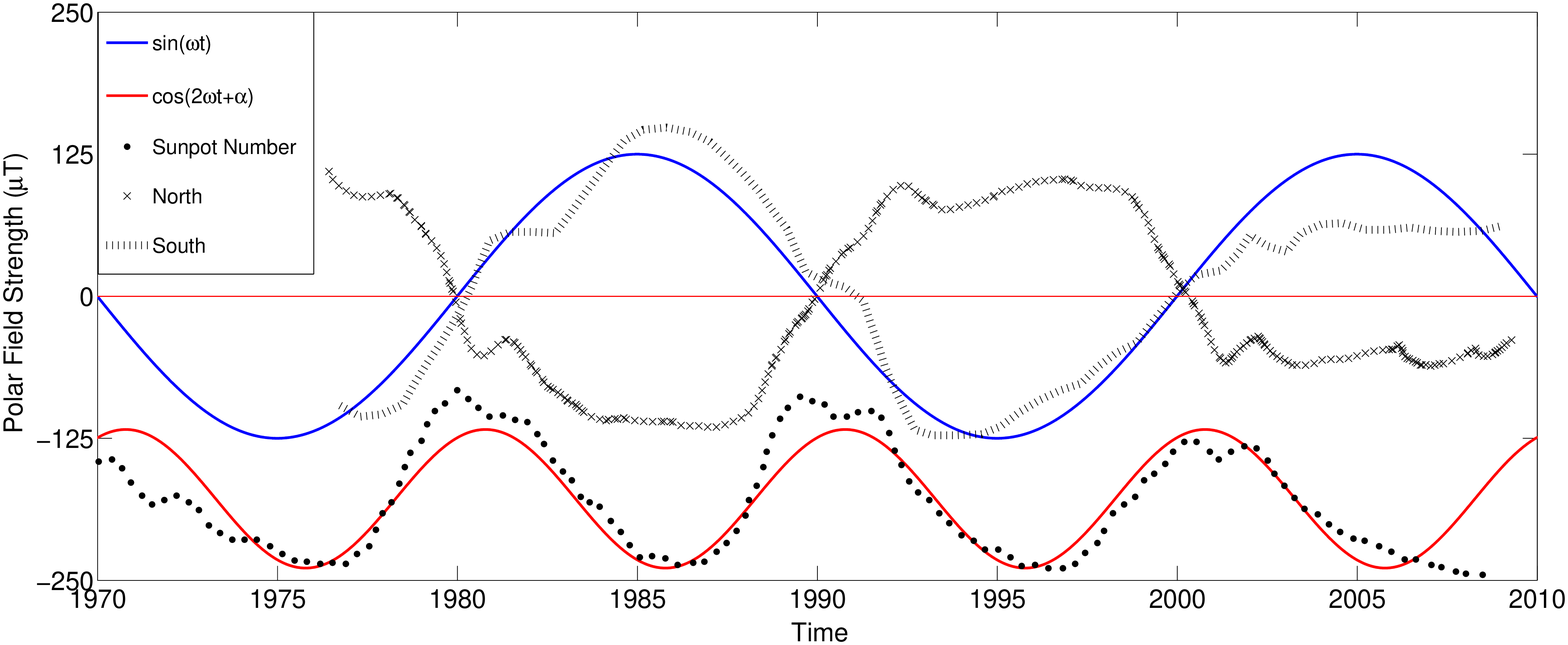}
\caption{\small Observed (black) and simulated number of sunspots and the polar field (red and blue).   
\label{spot}}
\end{figure}

As shown in Fig.$\ref{spot}$,  the poloidal field e.g., the south field was at a minimum amplitude in around 1980 (solar cycle $n=21$), then it increased with time and  peaked at 1985-1987(with $p^{-}_{n}\approx max$). Later it  reached the minimum again in around 1990.

The twisting of the south field of solar cycle 21 actually started at around  1980; and peaked in 1990 ($T^{-}_{n}\approx max$),  as shown in blue curve of Fig.$\ref{spot}$.  The phase of the toroidal field lags the poloidal field for approximately one fourth solar cycle ($\pi/2$), which is expected by the RLC circuit  as discussed in Section~2.


The peak of toroidal field, $T^{-}_{n}\approx max$ in 1990 (red in Fig.$\ref{disk}$B)  and 
the new toroidal field of next solar cycle 22 with opposite polarity, $T^{+}_{n+1}$, (blue in Fig.$\ref{disk}$B) become closer and closer with stretching of field lines through differential rotation, which provide site of magnetic reconnection responsible for solar activity and peak of sunspot in around 1990, as shown in Fig.$\ref{spot}$. 

In fact, as soon as $T^{-}_{n}$ first emerged in around 1980, it formed a site of magnetic reconnection with 
toroidal field of prior solar cycle,  $T^{+}_{n-1}$, which was  responsible for energy release and the peak of sunspots in around 1980.   

Therefore, in every solar cycle, the maximum  energy release is at the maximum $B_{\phi}$ corresponding to a  phase of   $\cos\omega t=1$ (recall $B_{\phi}\propto \cos\omega t$). E.g.,  for toroidal field, $T^{-}_{n}$, the first maximum energy release occurred when $T^{-}_{n}$ was as new toroidal field interacting with prior one,  $T^{+}_{n-1}$, and the second maximum of energy release happened when  $T^{-}_{n}$ as a old field interacting with the toroidal field of next solar cycle, 
 $T^{+}_{n+1}$. 
As  shown in  Fig.$\ref{disk}$, the toroidal field, $T^{+}_{n-1}$ being twisted for longer time is stronger in strength of magnetic field but weaker in flux  than  that of  $T^{-}_{n}$.







%

The large-scale toroidal field, $T^{+}_{n-1}$ of prior solar cycle and the current one, $T^{-}_{n}$,
 form X-point configuration, Fig.$\ref{disk}$ and  Fig.$\ref{rec}$a, which provide sites for magnetic reconnection. Can such reconnection explain the evolution of sunspots and  be rapid enough to account for the fast release of stored magnetic energy?


In the  reconnection of X-point configuration,  magnetic field lines are frozen into the charged particles, both ions and electrons, whose motion are  determined by ${\bf E}\times {\bf B}$, where ${\bf B}$ is $B_{n}$ and $B_{n-1}$ and $E=J_r/\sigma$ as shown in Fig.$\ref{rec}$abd.   Such a drift of  plasma (Hall current),  as shown in red lines of Fig.$\ref{rec}$bd, form ``magnetic nozzle".



To explain process of short timescales, \cite{Drake97, Shay99} has suggested that the energy release is instead mediated by electrons in waves called ``whistlers",  moving much faster for a given perturbation of the magnetic field with their smaller mass. And as the ``nozzle" becomes narrower, both the whistler velocity and associated plasma velocity increase. 

Such a fast reconnection mediated by whistler waves is supported by the finding  of  quadrupole structure in the  reconnection of magnetosphere\citep{DM01}.
  

 
 Hall effect in reconnection and the generation of magnetic field of quadrupole pattern is also discussed\citep{ZY09}.

 The current, $J_r$, in the center of Fig.$\ref{rec}$bd (perpendicular to the plane)  is driven by the radial emf, $E_r^{\prime}=E_R-E_r$. Such an out-plane  emf and the in plane magnetic field,  $T_{n-1}$  and $T_{n}$,  together drive the  plasma motion (Hall current) as shown in red arrows in Fig.$\ref{rec}$bd. Such an in-plane Hall current results in  quadrupole pattern of magnetic field in Fig.$\ref{rec}$bd in photosphere of the Sun.

As $T_{n-1}$ being twisted for longer time is stronger in strength of magnetic field than that of 
 $T_{n}$, the quadrupole pattern of magnetic field has stronger dipole field near the field line $T_{n-1}$ compared with that of $T_{n}$, as shown in Fig.$\ref{rec}$bd.

  In such a case,  a reversal of polarity of toroidal field corresponds to the reversal of the radial  emf as shown in  Eq.$\ref{ohm2}$.  Accordingly, the  polarity of Fig.$\ref{rec}$b and Fig.$\ref{rec}$d 
  change in the first half and second half of a solar cycle respectively, which  explains the reversal of polarity in the sunspots in each solar cycle, as shown in  http://solarscience.msfc.nasa.gov/images/magbfly.jpg.




The large sunspot pairs,  so called bipolar magnetic regions appear tilt with respect to the E-W direction, in which the leading sunspot, relative to the solar rotation,  is located at a lower latitude than the trailing sunspot, the pattern of which is known as Joy's law.  
This can be explained by the X-point configuration of twisted field which moves closer and closer to the equator in the twisting,  as shown in Fig.$\ref{disk}$ and Fig.$\ref{rec}$a.

Therefore, both the change of  polarity and configuration of sunspots can be interpreted by the   quadrupole structure in the  reconnection of magnetosphere\citep{DM01}, which is located between the prior and current toroidal field. 
 The observation  of flare ribbons  favors reconnection with preexisting field below the corona\citep{P17}, which can be seen as reconnection of such X-points above such sunspots. 
 Apparently, with the twisting of the prior and current cycle of field lines,  the location of X-points varies so that  the foot-point of such reconnection also moves.    
 
 

With a velocity of plasma of $\sim 10m/s$,  a length scale of $\sim 10$Mm and the magnetic diffusivity of Table~1,  
magnetic Reynolds number is $R_m=vL/\eta\sim 10^9$. 
Such an extremely large magnetic Reynolds number 
in this turbulent environment 
resulting in the efficient generation of magnetic fields on extremely small scales\citep{Jones10}, which explains activities, 
 e.g.,  nano-flares  to  coronal mass ejections.

\section{Modified Faraday disk vs Cowing's Law}
The  induction equation with constant conductivity is read, 
\begin{equation}{\label{Cow1}}
\frac{\partial {\bf B}}{\partial t}={\bf \nabla}\times({\bf u}\times {\bf B})+ \eta{\nabla^2}{\bf B}
\end{equation} 
In the case of  axisymmetric  magnetic field and  flow, a simpler decomposition is
\begin{equation}{\label{Cow2}}
{\bf B}=B\hat{\phi}+{\bf B}_z,   {\bf u}=s\Omega\hat{\phi}+{\bf u}_z
\end{equation} 
where $s=r\sin\theta$. The induction equation becomes\citep{Jones10},  
\begin{equation}{\label{Cow3}}
\frac{\partial {\bf B}}{\partial t}+s({\bf u}_z \cdot \nabla)(\frac{B}{s})=\eta(\nabla^2-\frac{1}{s^2})B+s{\bf B}_z \cdot \nabla\Omega
\end{equation} 
This equation reveals  some important aspects of the dynamo process. The advection term, $({\bf u}_z \cdot \nabla)$, and the diffusion term, $(\nabla^2-\frac{1}{s^2})$, cannot create magnetic field. Toroidal field can be generated from poloidal field via the term, $s{\bf B}_z\cdot\nabla\Omega$. In the case of the sun gradients of
angular velocity are both along radial and latitude with the latter much stronger than that if the former \citep{Tom03},  poloidal field is thus  stretched out by differential rotation of the latitude (as mentioned in Section 2) to generate toroidal  field. 

However, the poloidal field of Eq.$\ref{Cow3}$ has no source term, so it will just decay unless a mechanism can  be found  to sustain it\citep{Jones10}. This is the argument of Cowling's antidynamo theorem.  




In comparison,  Eq.$\ref{ohm2}$ represent induction on a modified Faraday disk, in which the term denoting stretched field lines by differential rotation is added to the last term at right hand side of   Eq.$\ref{ohm2}$ (the usual Faraday disk don't have such a term). In such a case $B_{\phi}$ is lagged for $\pi/2$ with respect to $B_z$ due to stretching of field lines through differential rotation. 

Consequently, there is no $B_{\phi}$ component that is in phase with component $B_z$ in Eq.$\ref{ohm2}$, so that the left hand side of Eq.$\ref{ohm2}$ corresponds to $\partial B/\partial t=\partial B_z/\partial t$, rather than two components in the first term at left hand side of Eq.$\ref{Cow3}$.

Most importantly, there is a term  extracting rotation energy of the Sun to generate a radial emf as shown in Eq.$\ref{Er}$, which is also displayed the second term at right hand side of Eq.$\ref{ohm2}$. It is this term that works as source to the poloidal field, and then converse to toroidal field which makes a self-sustained oscillation and thus avoided Cowling's antidynamo theorem.   In other words, Eq.$\ref{Cow3}$ is not self-sustain because it short of such a source term.


\section{Discussion}

As discussed above, the dynamo model base on   modified Faraday disk not only avoids Cowling's antidynamo theorem, so that an oscillation  of 22 years well account for solar cycle;  but also  provides site and current to trigger rapid magnetic reconnection with polarity change consisting with the behavior of sunspots.

 Notice that  magnetic diffusivity of Table~1 is obtained  
under Eq.$\ref{tan}$, which is consistent with  the oscillation period of Eq.$\ref{22c}$ calculated by
parameters $u_z$, $\kappa$, $S$, $r_1$, and $\mu$, $\xi$ of the Sun.

\begin{figure}
\includegraphics[width=0.45\textwidth]{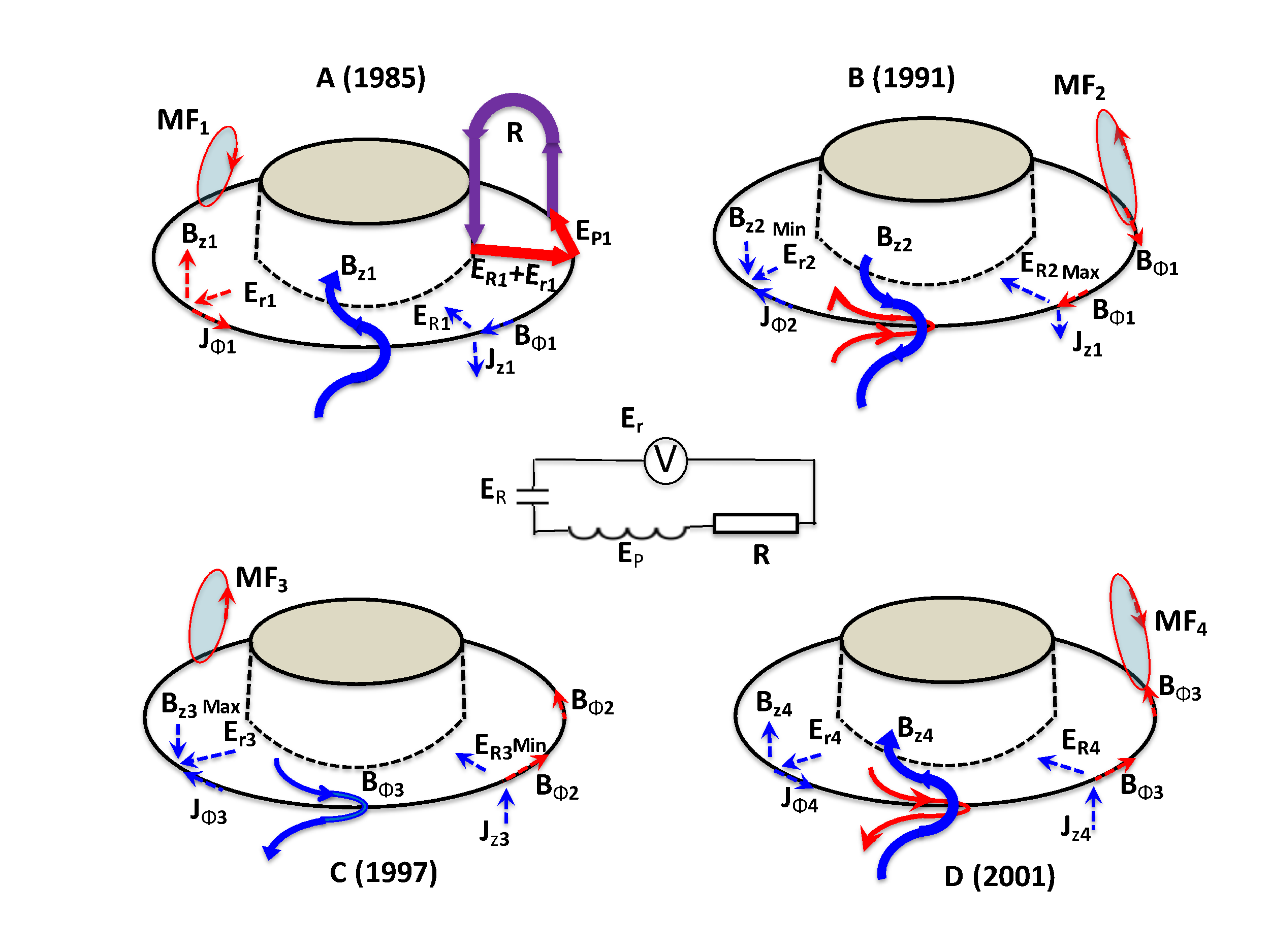}
\caption{\small A schematic show of the RLC circuit in the Faraday disk.   
The out and inner circles denote the radii   $0.95R_{\odot}$ and  $0.74R_{\odot}$ respectively in the equatorial plane of the Sun.
The ellipses in the four panels (A,B,C,D) represent the meridian flows (e.g., MF1 and MF2). 
The  red and blue curves denote toroidal field produced by the prior solar cycle and current cycle respectively.  
The closed lines in panel A  form a unit RLC circuit, with $E_r$ and $E_R$ in the radial plane of the disc,   $E_P$ along the disk ring, and the rest of the lines denote the resistance of the circuit. 
The sum of infinite number of such circuits  around the axis of the disk for 360 degrees gives the equivalent circuit in the middle. 
\label{disk}}
\end{figure}

\begin{figure}
\includegraphics[width=0.5\textwidth]{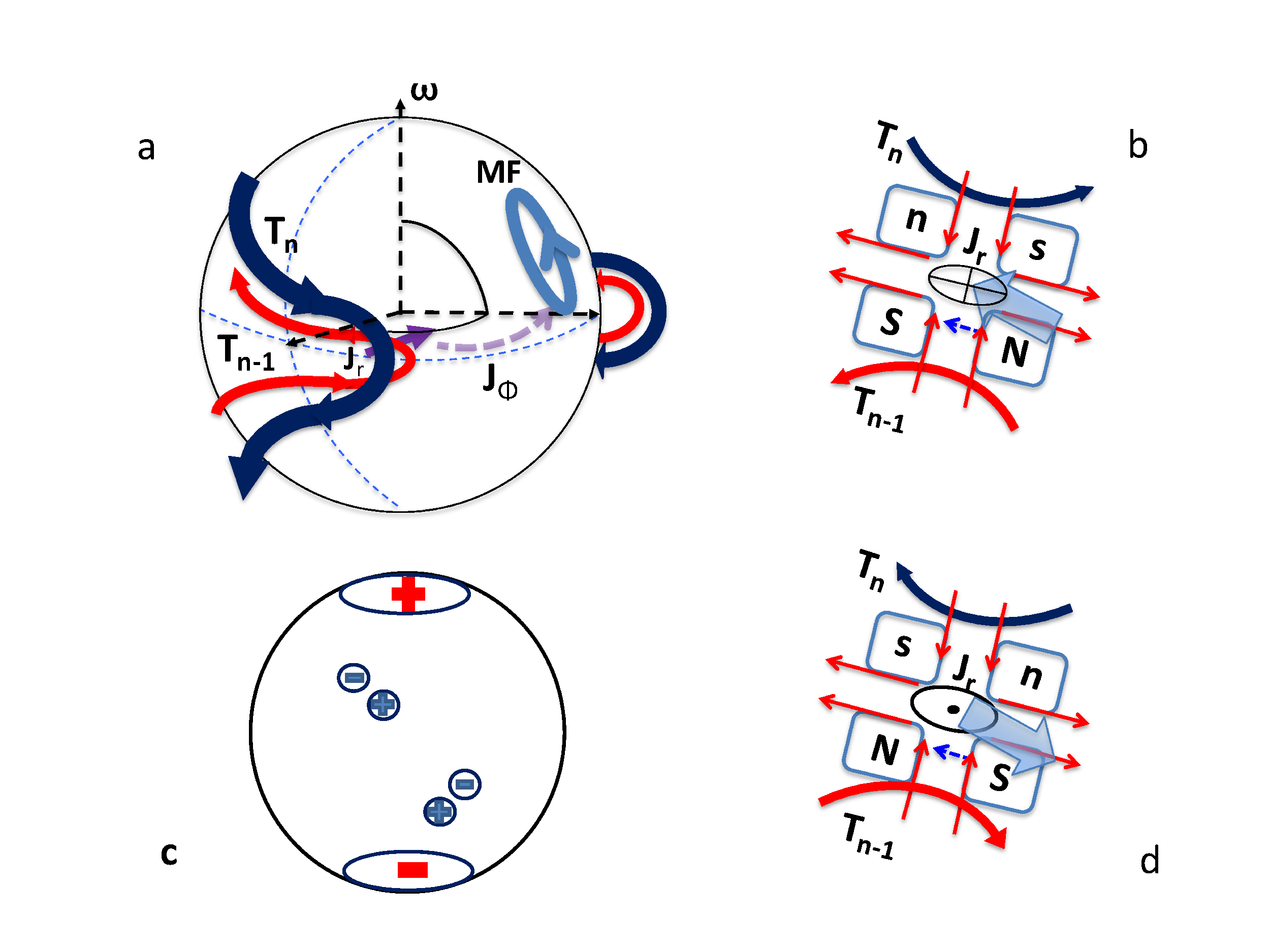}
\caption{\small A schematic show of twisted field and resultant reconnection configurations.   Panel a exhibits the configuration of toroidal field of prior and current solar cycle, which provides site of magnetic reconnection.
 Panel b   displays reconnection between $T_{n-1}$  and $T_{n}$ with current, $J_r$,  driven by the radial emf of the RLC circuit. Plasma motion (Hall current) as shown in red arrows  results in  quadrupole pattern of magnetic field.   Panel d  is the same as panel b but with opposite polarity. 
Panel c shows observational polarity of polar field and polarity of sunspots. 
\label{rec}}
\end{figure}

As discussed above, buoyancy of field lines is not necessary in the generation of sunspots, while  
considering buoyancy of the head of the twisted $B_{\phi}$,  as shown Fig.$\ref{rec}$a, and with current corresponding to purple curve of Fig.$\ref{disk}$a  reconnection  resembling the prominence is expected. 
Moreover, the polarity of such a  morphology changes with the polarity of  $B_{\phi}$ in each solar cycle,  which can be tested by observations. 



As shown in Fig.$\ref{disk}$ and Fig.$\ref{rec}$a , new model implies that the toroidal field at the northern and southern hemisphere of the Sun is  antisymmetric, which is consistent with observations. 

 The sum of toroidal field at different latitudes corresponds to an eddy current as shown in Fig.$\ref{disk}$a-d, which explains the meridian circulation  observed.  Moreover, the meridian circulation  of Fig.$\ref{disk}$a-d are affected by the change of polarity of the toroidal field in each solar cycle. However this don't mean that meridian circulation change from poleward to equator with the change of polarity of toroidal field. because the eddy current can change  sign by switch to positive or negative charge under the same velocity of meridian flow.      
 This  can be tested by further observation.

\begin{table}
\begin{center}
\caption{\bf Estimated parameters of the RLC circuit }
\begin{tabular}{ccccccccccc}
\hline
\hline
  $\epsilon (F/m)$ & $\mu (N/A^2)$ & $\sigma (S/m)$ & $\eta (m^2/s)$ & $C (F)$ & $L (H)$\\
  $1\times 10^3$ & $4\pi10^{-7}$ & $7\times10^6$ & $0.1$ & $2\times 10^{13}$ & $5\times 10^{2}$ \\
\hline 
\end{tabular}
\end{center}
\end{table}




\end{document}